\documentclass[aps,prl,reprint,amsmath,amssymb,floatfix,nofootinbib]{revtex4-2}

\usepackage[T1]{fontenc}
\usepackage[utf8]{inputenc}

\usepackage{bm}
\usepackage{mathtools}
\usepackage{graphicx}
\usepackage{xcolor}
\usepackage{bbm}
\usepackage{booktabs}

\usepackage{hyperref}
\hypersetup{
  colorlinks=true,
  linkcolor=blue!50!black,
  citecolor=blue!50!black,
  urlcolor=blue!50!black
}

\newcommand{\head}[1]{\medskip\noindent\textit{#1.--}}

\begin{document}

\title{Staying positive: bounds for non-Gaussian noise\\ in Schwinger-Keldysh effective field theory}

\author{Andrea Amoretti}
\author{Daniel K. Brattan}
\affiliation{Dipartimento di Fisica, Universit\'{a} di Genova, via Dodecaneso 33, I-16146, Genova, Italy.}
\affiliation{I.N.F.N. - Sezione di Genova, via Dodecaneso 33, I-16146, Genova, Italy.}

\date{\today}

\begin{abstract}
{\noindent Modern effective field theories of dissipative hydrodynamics, built on the Schwinger-Keldysh (SK) formalism, include thermal noise that is not Gaussian. The resultant higher-order noise vertices are constrained by symmetries, but otherwise treated as free parameters. We show explicitly that symmetry constraints alone do not guarantee these vertices yield a genuine probability distribution. Requiring a probability interpretation yields sharp bounds between cumulants that are invisible to the symmetry analysis. The bounds are optimal: their boundaries are attained by explicit positive distributions, and constructive completions are provided for both negative- and positive-kurtosis noise data. This gives closed-form nonlinear response and an exact sampling algorithm, whose Monte Carlo implementation reproduces the exact cumulants and nonlinear fluctuation-dissipation relations. Confronted with microscopic results, the framework places a sharp lower bound on the time step required for a positive stochastic completion of a quartic model of the Brownian particle used in the literature, while holographically computed $SU(2)$ noise vertices pass a nontrivial continuum fourth-order positivity test.}
\end{abstract}

\maketitle

\head{Introduction}%
Thermal fluctuations of conserved densities are crucial for explaining physically observed phenomena such as long-time tails and fluctuation corrections to transport. For example, non-Gaussian fluctuations of conserved charges are among the principal proposed signatures of the QCD critical point \cite{Kovtun2012,AnBasarStephanovYee2021,SogabeYin2022}. The modern framework that organizes them is the Schwinger-Keldysh (SK) effective field theory of dissipative hydrodynamics \cite{CrossleyGloriosoLiu2017,GloriosoCrossleyLiu2017,HaehlLoganayagamRangamani2018,GloriosoLiu2018,JensenMarjiehPinzaniYarom2018}, formulated on a doubled closed-time-path contour \cite{Schwinger1961,Keldysh1965}. For our purposes only one structural fact will be needed: in this formalism each conserved current comes with an auxiliary difference source $B_{a}$ that couples linearly to the fluctuating (noise) part of the current.

Concretely, let $\eta$ denote the noise accumulated in one spacetime cell - the region over which hydrodynamic and transport variables are coarse-grained to constancy - and let $b$ be the value of $B_{a}$ in that cell. If the SK noise sector admits a classical stochastic representation, its generating function must be the characteristic function of the accumulated noise,
\begin{equation}
\label{eq:Z}
Z(b)=\left\langle e^{i b\eta}\right\rangle =\int \mathrm{d}\eta\,P(\eta)\,e^{i b\eta},
\; \; i S(b)\equiv\log Z(b) \; ,
\end{equation}
with $P(\eta) \geq 0$. Any given SK action defines a candidate characteristic function $Z(b)=\exp[i S(b)]$, but whether it actually corresponds to a positive probability must be confirmed.

For a cell of spacetime volume $V_{\mathrm{cell}}$, we define the cumulants of \eqref{eq:Z} to be $\kappa_{n} = (-i)^n \partial_{b}^{n} \log Z(b) |_{b=0}$. For example, the statistical variance is $\kappa_{2}$. When the noise is local enough that it can be written as the sum of independent noise contributions in disjoint sub-cells of any given cell of volume $V_{\mathrm{cell}}$, then we can write $\kappa_{n} = c_{n}V_{\mathrm{cell}}$ where $c_{n}$ is the $n^{\mathrm{th}}$ cumulant density.

Thermal equilibrium is imposed on $S$ in \eqref{eq:Z} by the dynamical Kubo-Martin-Schwinger (KMS) symmetry \cite{Kubo1957,MartinSchwinger1959,HHW1967}, which enforces fluctuation-dissipation relations order by order \cite{CrossleyGloriosoLiu2017,GloriosoCrossleyLiu2017,GloriosoLiu2018,JensenMarjiehPinzaniYarom2018}. Beyond quadratic order in $B_{a}$ the theory contains genuinely non-Gaussian noise, which arises in microscopic Brownian models \cite{LinBuLei2024} and in holographic SK constructions \cite{BuSunZhang2022,BuZhang2021,BuDemircikLublinsky2021}, and is used phenomenologically as stochastic transport coefficients in critical fluctuations, hydrodynamic three-point functions, and long-time tails \cite{JainKovtun2022,AnBasarStephanovYee2021,SogabeYin2022,AnBasarStephanovYee2022,AbbasiRischke2025,AnBasarStephanov2026,BasarSong2026}.

These developments raise a question that the SK axioms do not answer: \emph{when does the function $Z(b)$ defined by an SK action actually come from a probability distribution $P(\eta)\ge0$, as in Eq.~\eqref{eq:Z}?} A classical theorem of Bochner \cite{Bochner1933} states that this happens if and only if $Z$ is a normalised, continuous, \emph{positive-definite} function, $\sum_{m,n}\alpha_m^*\alpha_n\,Z(\beta_n-\beta_m)\ge0$ for all finite collections of real $\beta_n$ and complex $\alpha_n$, with a field-theoretic version due to Minlos \cite{Minlos1959,GelfandVilenkin1964}.  This condition is not implied by the standard local SK constraints, including reality, normalization, dynamical KMS symmetry, and the commonly imposed condition $\mathrm{Im}[S] \geq 0$ \cite{GloriosoLiu2018,JensenMarjiehPinzaniYarom2018}. Thus SK-consistent pure-noise vertices need not be stochastically realisable, and requiring a positive stochastic completion imposes independent constraints on EFT coefficients. When positive definiteness fails, Fourier inversion does not define a positive probability measure; when the inverse exists as an ordinary
function, it must develop negative regions. Such a signed or generalized quasi-distribution may reproduce correlators algebraically,
but it cannot serve as an ordinary positive noise law for direct Langevin sampling.

In this Letter we show that this failure occurs for realistic truncated data; we derive sharp, quantitative bounds that genuine positivity imposes on the non-Gaussian vertices and that are invisible to order-by-order KMS analysis. We prove the bounds optimal by constructing exact positive distributions that attain the relevant bounds and providing closed-form completions for broad classes of admissible truncated data, and we confront the resulting constraints with published microscopic vertices. 

If the noise varies from one spacetime-cell to another then for an entire system, \eqref{eq:Z} becomes
	\begin{eqnarray}
		\label{Eq:Celldependentnoise}
		Z[B_{a}] = \int D\eta\,P(\eta)\,e^{i \int B_{a}(x) \eta(x)} \; . 
	\end{eqnarray}
Throughout, for simplicity, we work with local, zeroth-derivative noise (LZDN) in the classical ($\hbar\to0$) limit. The pure-noise sector then contains no derivatives of $B_a$, and \eqref{Eq:Celldependentnoise} factorizes into cell-wise laws of the form \eqref{eq:Z}. Derivative-dependent noise instead requires the functional form \eqref{Eq:Celldependentnoise} , with frequency- and momentum-dependent cumulant kernels \cite{AmorettiBrattanRongen2026,AmorettiAnselmiBrattanBranchCuts}. The hydrodynamic state in LZDN may nevertheless vary slowly in spacetime, allowing genuine fluid flows, provided its local values are fixed when defining the noise law of each coarse-graining cell. 

All proofs, the tensor generalizations, and the extensions are collected in the Supplemental Material (SM) \cite{SM}.

\head{A truncation that is not noise}%
At zero external force, KMS and reality allow a quartic action of the form
\begin{equation}
i S(b)\;=\;-\frac{\kappa_{2}}{2}\,b^2\; + \frac{\kappa_{4}}{4!} b^4 ,
\label{eq:quartic}
\end{equation}
where $\kappa_2$ is the Gaussian noise strength (for charge diffusion, the Landau-Lifshitz value $\kappa_{2}=2T\sigma V_{\mathrm{cell}}$ with $T$ the temperature and $\sigma$ the conductivity \cite{LandauLifshitzFluid,Kovtun2012}) and $\kappa_{4}$ is the fourth cumulant. Data of exactly this form are produced by microscopic matching \cite{LinBuLei2024}. 

\begin{figure}[t]
\includegraphics[width=\columnwidth]{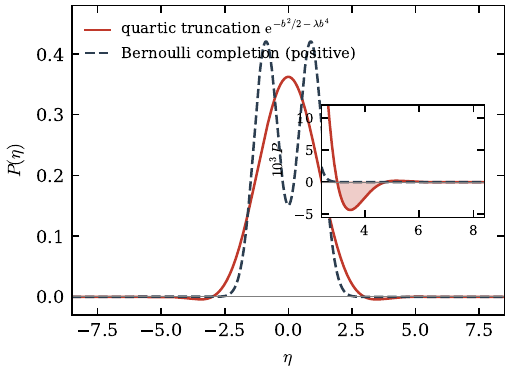}
\caption{The distribution reconstructed from the KMS-consistent quartic action~\eqref{eq:quartic} with $\kappa_{4}=-1.2$ (solid) is negative for $3.0\lesssim|\eta|\lesssim4.5$ (inset, shaded). It is a quasi-probability and cannot be sampled directly as an
ordinary positive noise distribution. The dashed curve is the Bernoulli completion described below Eq.~\eqref{eq:k4bound}, namely random kicks $\pm q$ blurred by a Gaussian, which matches the variance and fourth cumulant of the truncation exactly and is a genuine probability density.}
\label{fig:neg}
\end{figure}

Take $\kappa_{2}=1$, $\kappa_{4}=-1.2$ in \eqref{eq:quartic}. Inverting the Fourier transform in Eq.~\eqref{eq:Z} gives a function $P(\eta)$ that is \emph{negative} on an entire interval, $3.0\lesssim|\eta|\lesssim4.5$, with minimum $\simeq-4.3\times10^{-3}$ [Fig.~\ref{fig:neg}]. This negativity cannot be repaired by tuning - a theorem of Marcinkiewicz \cite{Lukacs1970,RajagopalSudarshan1974} states that the exponential of a polynomial of degree greater than two is \emph{never} a characteristic function. Therefore, no finite polynomial exponent of degree greater than two can define an exact characteristic function on the full real axis. Such polynomial SK actions may consistently be used as low-order EFT truncations, but not as exact positive noise laws. The relevant question to us is therefore whether the finite set of cumulants encoded by the truncation admits at least one positive, KMS-compatible completion.

\head{KMS as local detailed balance}%
To formulate KMS in the noise sector, consider one coarse-graining cell and displace the system by a thermodynamic force $F$, conjugate to $\eta$, which is constant across the cell but may otherwise vary from one cell to another. As an example, for charge diffusion the force would be applied electric field in units of the temperature. In the classical limit the dynamical KMS transformation acts on the difference source simply as $b\to-b+i F$ \cite{GloriosoCrossleyLiu2017,GloriosoLiu2018}.  Since $b \eta$ is dimensionless and KMS acts as $b\to -b+iF$, $F$ has the same units as $b$, so that $F\eta$ is dimensionless.

After displacing by a force, the noise distribution becomes $F$-dependent, $P_F(\eta)$, with characteristic function $Z_F(b)$. Under the regularity assumptions stated in the SM, KMS invariance is then equivalent to a detailed-balance property of the distribution itself:
\begin{equation}
Z_F(b)=Z_F(-b+i F)
\;\Longleftrightarrow\;
\frac{P_F(\eta)}{P_F(-\eta)}=e^{\,F\eta}.
\label{eq:DB}
\end{equation}
The right-hand side has a transparent physical meaning, a noise kick $\eta$ along the force is more likely than the reversed kick $-\eta$ by the exponential of the work $F\eta$. This is a Crooks-type relation \cite{Crooks1999,MullinsHippertNoronha2025} holding here at the level of the noise law itself. The proof uses only uniqueness of Fourier transforms and holds for quasi-probabilities too. Crucially, positivity is logically \emph{independent} of KMS, which is precisely why it must be imposed separately. 

Whenever the required exponential moment is finite, every positive solution of Eq.~\eqref{eq:DB} can be decomposed into:
\begin{subequations}
\label{eq:master}
\begin{align}
& P_F(\eta) = \frac{e^{\,F\eta/2}\,Q_F(\eta)}{\int d\tilde{\eta} \; e^{\,F\tilde{\eta}/2}\,Q_F(\tilde{\eta})} \\
 \;\;\Longrightarrow\;\; &
i S(b,F)=\mathcal C_F\!\big(\tfrac F2+i b\big)-\mathcal C_F\!\big(\tfrac F2\big),
\end{align}
\end{subequations}
where $Q_F(\eta)\propto e^{-F\eta/2}P_F(\eta)$ is the positive
symmetric seed, $Q_F(\eta)=Q_F(-\eta)$, and
$\mathcal C_F(x)=\log\int\mathrm{d}\eta\,Q_F(\eta)e^{x\eta}$
is its cumulant-generating function. Equation~\eqref{eq:master} is an exact, to all orders in $b$ within the LZDN sector, KMS-invariant action. The split is clear, KMS fixes the forward/backward bias while $Q_{F}$ encodes the symmetric microscopic behaviour.

Using the definition of the cumulants, Eq.~\eqref{eq:master} leads to the following recursion relations
\begin{eqnarray}
\label{eq:gradFDT}
\kappa_{n+1}(F)\;=\;2\,\frac{\mathrm{d} \kappa_n(F)}{\mathrm{d} F} - \left[ 2  \partial_{F}  \partial_{x}^{n} \mathcal{C}_{F} \right]_{x=\frac{F}{2}} \; .
\end{eqnarray}
In what follows, for pedagogical simplicity, in examples we will generally consider a force-independent seed $Q_{F}=Q_{0}$. Thus the last term in \eqref{eq:gradFDT} vanishes, giving $\kappa_{n+1} = 2 d\kappa_{n}/dF$ at finite force. Within this class, non-Gaussian noise and nonlinear response are rigidly tied together. Below, the positivity results do not require this restriction unless otherwise stated.

Equation \eqref{eq:DB} is the familiar Gallavotti-Cohen/Lebowitz-Spohn fluctuation symmetry \cite{GallavottiCohen1995,LebowitzSpohn1999,AndrieuxGaspard2007,TobiskaNazarov2005,SaitoUtsumi2008,EspositoHarbolaMukamel2009}, while equation \eqref{eq:master} is an Esscher tilt \cite{Esscher1932}. Here we identify this structure directly with dynamical KMS in the LZDN sector and separately impose positivity of the noise law. This is stronger than positivity of $\mathrm{Im}[S] \geq 0$, as assumed in \cite{MullinsHippertNoronha2025}.

\begin{figure}[t]
\includegraphics[width=\columnwidth]{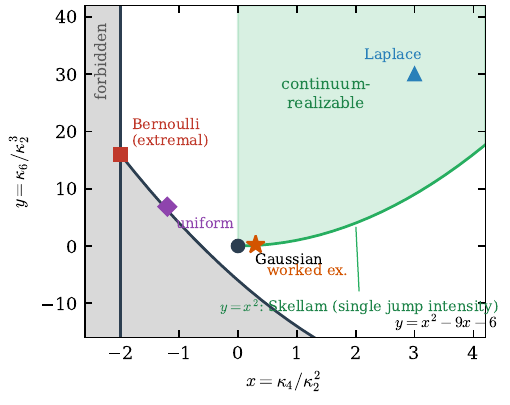}
\caption{Allowed region for the fourth and sixth noise cumulants at $F=0$, normalized by the variance. For a finite cell size, positivity requires $x\ge-2$ and $y\ge x^2-9x-6$ [Eqs.~\eqref{eq:k4bound} and~\eqref{eq:k6bound}]. The two-point (Bernoulli) law sits at the extremal corner. In the local continuum limit (green) the zero-force Stieltjes hierarchy further requires $x\ge0$ and $y\ge x^2$, the boundary $y=x^2$ corresponding to kicks of a single intensity. Symbols mark the exactly solvable completions of the SM while the star is the worked example of Table~\ref{tab:MC}.}
\label{fig:region}
\end{figure}

\head{What positivity forbids}%
The Bochner theorem discussed above, using the definition of the cumulants and \eqref{eq:master}, leads to bounds on combinations of the cumulants. For example, positivity of the centred moment matrix gives
\begin{equation}
\kappa_{4}(F)  \geq \frac{\kappa_{3}^2(F)}{\kappa_{2}(F)} - 2 \kappa_{2}^2(F)  \stackrel{F\rightarrow 0}{\longrightarrow} \kappa_4\ge-2\kappa_2^2
\label{eq:k4bound}
\end{equation}
where $\kappa_{3}(0)=0$ by detailed balance. Generally, at finite $F$, positivity bounds will involve odd cumulants, that otherwise disappear in the $F=0$ limit. The zero-force bound in \eqref{eq:k4bound} is saturated by a two-point law. Similarly, at $F=0$ the sixth-order cumulant is first constrained by (SM, Sec. SV)
\begin{equation}
\kappa_2\,\kappa_6\;\ge\;\kappa_4^2-9\,\kappa_2^2\kappa_4-6\,\kappa_2^4 .
\label{eq:k6bound}
\end{equation}
At the extremal corner $\kappa_{4}=-2\kappa_{2}^2$, Eq. \eqref{eq:k6bound} is also saturated by a two-point law [Fig.~\ref{fig:region}] given by kicks $\pm q$, with $q = (-\kappa_4/2)^{1/4}$ such that $G = 0$. Positive-kurtosis completions are given in the SM.

\head{The continuum limit}%
In the local white-noise limit, valid when microscopic noise-correlation scales are short compared with the hydrodynamic coarse-graining scale, we take disjoint cells to have independent increments and cumulants to be extensive in cell volume. These assumptions force the single-cell distribution to be infinitely divisible, and the L\'evy-Khintchine theorem \cite{Sato1999} then implies a Gaussian component of variance density $G$, superposed with independently scattered jumps governed by a positive L\'evy measure $\nu(\mathrm{d} q)$. The cumulant densities, provided the relevant moments are finite, are
\begin{eqnarray}
c_{k}(F)
& = G(F) \delta_{k,2} +
\int q^{k}\,\nu_{F}(\mathrm{d} q) \; ,
\end{eqnarray}
where $k \geq 2$, $\delta_{k,2}=1$ when $k=2$, zero otherwise.

To place the Gaussian and jump contributions in a common moment representation, define a positive measure $\widetilde{\sigma}_{F}$ on $\mathbbm{R}$ by
\begin{eqnarray}
\widetilde{\sigma}_{F}(\mathrm{d} q)
&:=&    G(F) \delta_{0}(\mathrm{d} q) + q^2 \nu_{F}(\mathrm{d} q)
\end{eqnarray}
The first term of weight $G$ represents the Gaussian component, while the other term describes the jump intensities. It then
follows that
\begin{equation}
c_{k+2}(F) = \int_{-\infty}^\infty q^k\,\widetilde{\sigma}_{F}(\mathrm{d} q), \qquad k=0,1,2,\ldots
\label{eq:stieltjes}
\end{equation}
Thus $\left\{ c_{n+2}(F) \right\}$ form a Hamburger moment sequence. In particular,
\begin{eqnarray}
\label{eq:c4bound}
c_{2}(F) c_{4}(F) \geq c_{3}(F)^2 \; .
\end{eqnarray}
At $F=0$, symmetry removes the odd cumulants and the even sequence obeys the Stieltjes hierarchy discussed in the SM. In fact, for a force-independent seed, we can combine \eqref{eq:gradFDT} and \eqref{eq:c4bound} to give
\begin{equation}
   \label{Eq:CombinedBound}
    c_4(F)
    \geq
    \frac{4\left[\partial_F c_2(F)\right]^2}{c_2(F)} \, ,
\end{equation}
which constrains $c_{4}(F)$ in terms of how quickly $c_{2}(F)$ changes with $F$.

Two clear consequences stand out: firstly, within the ultralocal independent-increment class considered here, a negative fourth-cumulant density is excluded. A negative fourth cumulant may nevertheless occur at finite coarse-graining scale or outside this class, for example for colored or spatially correlated noise. Second, nonlinear stochastic effective theories generally contain several a priori independent higher-order noise and interaction coefficients \cite{JainKovtun2022,SogabeYin2022,AbbasiRischke2025,AmorettiAnselmiBrattanMaxwellCattaneo}. Whenever these coefficients can be identified with local noise cumulant densities, positivity imposes the additional relations above. The derivation assumes cell independence and cumulants proportional to the cell volume. These assumptions cease to be valid when the coarse-graining scale is not large compared with the correlation length, as can occur near a critical point. In that regime the continuum L\'evy/Stieltjes bounds do not apply as stated, although the finite-cell positivity constraints remain valid. The generalization to this case is an open problem.

\head{Monte Carlo}%
The kick picture is not just a proof device, it supplies ready-to-use noise models. For charge diffusion in $d$ dimensions, take kicks of fixed magnitude $q_0$ in uniformly random directions at rate $r$, plus a Gaussian remainder tuned so that the total variance matches Landau-Lifshitz, $G=2T\sigma-rq_0^2/d\ge0$. All higher cumulants then follow in closed form from a single special function (SM, Sec.~SVIII). The decisive property of this class is how it responds to a force. The half-force tilt of a kick distribution is again a kick distribution, with rates reweighted as
\begin{equation}
\nu_F(\mathrm{d} \vec{q})\;=\;e^{\,\vec{q} \cdot \vec{F}/2}\;\nu(\mathrm{d} \vec{q}).
\label{eq:tiltednu}
\end{equation}
The force does not deform the kicks, it makes kicks along the force more frequent in exact Crooks proportion. This realizes detailed balance microscopically, and it \emph{is} a sampling algorithm. Per cell one: (i) draws the Gaussian part, (ii) draws the number of kicks from a Poisson law with the tilted total rate and (iii) draws each kick direction from the tilted angular distribution. Every step is standard positive sampling, with no negative weights anywhere.

We verify the sampling construction by generating positive samples and confirming that their measured cumulants reproduce the exact analytic results. In particular, for $d=3$ with $T=1$, $\kappa_{2}=2T\sigma V_{\mathrm{cell}}=1$, $q_0=1$, $r=3/2$, $G=1/2$, so that half the variance is carried by kicks; the zero-force cumulant densities $c_2=1$, $c_4=3/10$, $c_6=3/14$ satisfy the Stieltjes inequalities. At force $F$ all cumulants are elementary closed forms (SM, Sec.~SIX), obeying Eq.~\eqref{eq:gradFDT} identically. Table~\ref{tab:MC} compares them at $F=1$ with $2\times10^7$ sampled cells. There is agreement within statistics for all four measured cumulants. Moreover, because the analytic values satisfy $\kappa_3=2\,\mathrm{d}\kappa_2/\mathrm{d} F$ identically, the agreement of the measured skewness provides a direct sampling check of the nonlinear FDT hierarchy, with the skewness of genuinely sampled noise tracking the force-susceptibility of its variance beyond linear response. Moreover, the sampled cumulants satisfy the finite-force continuum bound $c_2c_4\ge c_3^2$. For the fixed-seed completion used here, $c_3=2\partial_F c_2$, so this inequality is precisely equivalent to \eqref{Eq:CombinedBound}. The Monte Carlo therefore also provides a direct sampling check of the finite-force positivity bound for this completion.

One distinction is worth keeping sharp: the non-equilibrium, non-zero force response of this model, an effective conductivity $\kappa_1/F$ that grows exponentially once $q_0F\gtrsim1$, is a prediction of \emph{this} completion, while the parameter-free relations~\eqref{eq:gradFDT} are shared by \emph{every} fixed-seed positive noise and are falsifiable in any realization.

\head{Cumulants from current simulations}%
The cumulants are not abstract information and can be determined from repeated current simulations. Let $\mathcal C$ be a fixed coarse-graining cell.  For a chosen direction $\hat n_i$, the accumulated noise is
\begin{equation}
\eta_{\mathcal C}
=
\int_{\mathcal C} d^{d+1}x \;
\hat n_i \left[J^i(x)-J_{\rm hydro}^i(x)\right].
\end{equation}
Here $J^i$ denotes the full fluctuating current and $J_{\rm hydro}^i$ its deterministic hydrodynamic contribution. Repeating the same measurement gives the distribution $P(\eta_{\mathcal C})$ and hence its cumulants. For example
\begin{equation}
\kappa_2
=
\left\langle \eta_{\mathcal C}^2\right\rangle
-
\left\langle \eta_{\mathcal C}\right\rangle^2 .
\end{equation}
For a slab-like cell of thickness $\ell$ normal to a surface $\Sigma$, this is related to the transferred charge
\begin{equation}
Q=\int_{\Delta t}dt\int_\Sigma d\Sigma_i J^i
\end{equation}
by $\eta_{\mathcal C}\simeq \ell (Q-Q_{\rm hydro})$, connecting directly to full-counting-statistics measurements of charge transfer \cite{Gustavsson2006,Fricke2010}.  Varying the cell size then tests whether $\kappa_n\propto V_{\rm cell}$, as required to define the continuum cumulant densities $c_n$.

\begin{table}[t]
\caption{Monte Carlo cumulants at $F=1$ ($2\times10^7$ cells, errors from 20 independent batches) and $V_{\mathrm{cell}}=1$ versus the exact closed forms (SM, Sec.~SIX).}
\label{tab:MC}
\begin{ruledtabular}
\begin{tabular}{lccc}
 & exact & Monte Carlo & pull \\
\colrule
$\kappa_1$ & $0.506306$ & $0.506634\pm0.000217$ & $+1.5$ \\
$\kappa_2$ & $1.038062$ & $1.038208\pm0.000348$ & $+0.4$ \\
$\kappa_3$ & $0.154508$ & $0.154539\pm0.000923$ & $+0.0$ \\
$\kappa_4$ & $0.327223$ & $0.329269\pm0.001732$ & $+1.2$ \\
\end{tabular}
\end{ruledtabular}
\end{table}

\head{Confronting microscopic vertices}%
The bounds developed here bite on existing results, in both directions. Lin, Bu, and Lei \cite{LinBuLei2024} study the SK theory of a Brownian particle whose noise sector reads, in our notation, $i S=-Tb^2-\epsilon_1b^4$ per unit time, so that $\kappa_{2}=2T$ and $\kappa_{4}=- 4! \epsilon_1$, with $\epsilon_1>0$ required by stability of their Fokker-Planck equation. They represent the quartic vertex as a separate noise with vanishing variance and fourth moment $-24\epsilon_1/\delta t^3$ at time step $V_{\mathrm{cell}}=\delta t$, find its weight function to be nonpositive, and argue that this is unavoidable. The zero-force bound following from Eq.~\eqref{eq:k4bound}, $\kappa_4\ge -2\kappa_2^2$, shows that the negativity is a property of that splitting rather than of the total noise, and upgrades the observation to a quantitative statement: the combined single-step cumulants admit a positive completion if and only if the coarse-graining time obeys
\begin{equation}
    \delta t \ge \frac{3\epsilon_1}{T^2},
\end{equation}
a sharp lower bound saturated by the Bernoulli completion. Thus, for fixed $\epsilon_1>0$, no positive independent-increment continuum limit exists as $\delta t\to0$, consistently with the negative fourth-cumulant density $c_4=-24\epsilon_1$. A joint limit
$\delta t,\epsilon_1\to0$ can evade this obstruction only if $\epsilon_1$ vanishes at least as rapidly as required by the bound above.

The comparison runs the other way for genuinely holographic data. For SU(2) diffusion in AdS$_5$, Bu, Sun, and Zhang computed the quartic noise vertices analytically \cite{BuSunZhang2022}. Writing
\begin{equation*}
M^{\rm ab}=\sum_i B^{\rm a}_{ai}B^{\rm b}_{ai}
\end{equation*}
for the flavour-space Gram matrix of the spatial difference sources, their two pure-noise couplings, although individually of opposite sign, combine into
\begin{equation*}
i S_4=
\frac{\pi}{128}
\left[
(\operatorname{tr}M)^2-\operatorname{tr}(M^2)
\right].
\end{equation*}
To evaluate this form along an arbitrary normalized source direction, set
\begin{equation*}
B^{\rm a}_{ai}=b\,V^{\rm a}_i,
\qquad
\sum_{{\rm a},i}(V^{\rm a}_i)^2=1, \qquad
(M_V)^{{\rm ab}}=\sum_iV^{\rm a}_iV^{\rm b}_i.
\end{equation*}
Since $M_V$ is positive semidefinite and
$\operatorname{tr}M_V=1$, the directional fourth-cumulant density is
\begin{equation}
c_4(V)
=
\frac{3\pi}{16}
\left[
1-\operatorname{tr}(M_V^2)
\right]
\geq0.
\label{eq:holok4}
\end{equation}
Equality holds precisely for factorized rank-one directions $V^{\rm a}_i=v^{\rm a}e_i$. Generic nonfactorized directions give a
strictly positive fourth cumulant. Thus all scalar projections of the holographic quartic vertex pass the continuum fourth-cumulant positivity test nontrivially. Any extension of the same computation to sextic order must consequently satisfy $c_2(V)c_6(V)\geq c_4(V)^2$ along every normalized combined flavor-spatial direction $V^{\rm a}_i$. Details, and the caveats attached to the temporal components and to the quantum level of the holographic action, are given in the SM, Sec.~SX.

\head{Outlook}%
The emerging structure is simple. Classical KMS in the LZDN sector is detailed balance of the noise law, solved in general by an exponential tilt of a symmetric seed. A force-independent seed $Q_{F}=Q_{0}$ ties all nonlinear response to noise non-Gaussianity through Eq.~\eqref{eq:gradFDT}. Positivity is an independent axiom with quantitative teeth, comprising the Marcinkiewicz no-go, the cumulant bounds, and the Stieltjes hierarchy, none of which is visible to symmetry analysis. The confrontation with microscopic results shows both faces of the framework at work: a widely used stochastic truncation acquires a sharp bound on its regularization scale, and the scalar projections of a genuinely holographic vertex set pass the continuum fourth-cumulant positivity test.

Published pure-noise vertices arising in microscopic Brownian models \cite{LinBuLei2024}, holographic constructions
\cite{BuSunZhang2022}, loop calculations \cite{SogabeYin2022}, or phenomenological effective theories \cite{JainKovtun2022,AbbasiRischke2025,AnBasarStephanov2026} can be screened for positive stochastic realizability whenever they can be consistently identified with cell-wise cumulants in the LZDN sector. Whenever the test is passed, the tilted-kick
constructions provide explicit, exactly sampleable positive completions for that sector. In the same phenomenological context, positivity and maximum entropy have been used to constrain the freeze-out of hydrodynamic fluctuations into particle distributions \cite{PradeepStephanov2023}. The constraints derived here act one step earlier, on the noise that drives the fluctuations themselves. Beyond the LZDN sector, three extensions stand open: derivative-dependent noise, including frequency-dependent (colored) and spatially correlated noise, where KMS acts on time arguments; backgrounds that break time reversal, where detailed balance relates conjugate ensembles; and the quantum regime, where the natural analogue of positivity is complete positivity of the influence functional.

\begin{acknowledgments}
A.A. \& D.B. have received support from the project PRIN 2022A8CJP3 by the
Italian Ministry of University and Research (MUR). D.B. is currently funded by
PNRR GIOVANI RICERCATORI, CUP D33C25000470006.
\end{acknowledgments}

\clearpage
\onecolumngrid
\setcounter{section}{0}
\setcounter{secnumdepth}{1}
\setcounter{equation}{0}
\setcounter{table}{0}
\renewcommand{\theequation}{S\arabic{equation}}
\renewcommand{\thesection}{S\Roman{section}}
\renewcommand{\thetable}{S\Roman{table}}

\begin{center}
{\large\bfseries Supplemental Material for\\[3pt]
``Staying positive: bounds for non-Gaussian noise\\
in Schwinger-Keldysh effective field theory''}

\vspace{0.6em}
Andrea Amoretti and Daniel K. Brattan

\vspace{0.4em}
\today
\end{center}

\vspace{0.8em}

\noindent This Supplemental Material collects additional conventions, derivations, and constructions not included in the main text. Section~\ref{sec:conv} fixes the setting and the three nested notions of positivity; Sec.~\ref{sec:kms} derives the equivalence of dynamical KMS invariance and detailed balance, in the reduced convention and for a general time-reversal involution; Sec.~\ref{sec:tilt} shows that the half-force tilt is the general positive solution and maps out its domain; Sec.~\ref{sec:fdt} derives the master action and the gradient form of the nonlinear fluctuation--dissipation (FDT) hierarchy, including its converse; Sec.~\ref{sec:obst} derives the positivity bounds; Sec.~\ref{sec:cont} treats the continuum limit and the Hamburger structure of the cumulant densities; Sec.~\ref{sec:lib} presents the library of exact seeds; Sec.~\ref{sec:hydro} performs the tensorial matching to charge-diffusion and stress noise; Sec.~\ref{sec:mc} derives the tilted-kick sampling algorithm and documents the Monte Carlo; and Sec.~\ref{sec:micro} confronts the positivity constraints with the microscopic Brownian and holographic vertices discussed in the main text. Equations of the main text are referred to as (1), (2),~\dots; equations here carry the prefix~S.

\section{Setting and the three notions of positivity}
\label{sec:conv}

We work in a finite-dimensional real vector space of noise variables $\eta^A$, where the multi-index $A$ may collect spacetime-cell labels, internal indices, vector indices, or symmetric-tensor indices. The extension to field theory is discussed in Sec.~\ref{sec:cont}, and the main text simply suppresses $A$. Pairings for the cell are written $(J,\eta)=J_A\eta^A$ and $(\eta,B_{a})=\eta^AB_{aA}$. In the continuum of cells these pairings become, for example,
	\begin{displaymath}
		(J,\eta) = \int d^{d+1}x \; J_\mu(x) \eta^\mu(x) \; ,
	\end{displaymath}
where $\mu$ is a spacetime index. The SK difference source is $B_{a}$, and the thermodynamic force built from the physical ($r$-type) backgrounds is denoted schematically $F\equiv L_\beta B_r$. In the classical $(\hbar\to0)$ limit, and restricted to the local, zeroth-derivative pure-noise (LZDN) sector, the dynamical KMS transformation acts component-wise on the difference source as
\begin{equation}
(\Theta B_{a})_{A}=-B_{aA}+i F_{A},\qquad (\Theta F)_{A}=F_{A},
\label{eq:reducedKMS}
\end{equation}
which we call the reduced convention. Here ‘zeroth derivative’ refers to derivatives acting on the difference source $B_a$; the force $F\equiv{\cal L}_\beta B_r$ may itself contain derivatives of the physical $r$-type backgrounds. The general convention, including time reversal and tensor parities, appears in Sec.~\ref{sec:kms}. Given a (possibly only quasi-) probability density $P_F(\eta)$, the noise generating functional and action are
\begin{equation}
Z_F[B_{a}]=\int\!D\eta\;P_F(\eta)\,e^{i(\eta,B_{a})},\qquad
i S[B_{a},F]=\log Z_F[B_{a}].
\label{eq:Zdef}
\end{equation}

Three nested notions of positivity organize everything that follows. The weakest, \emph{SK consistency}, is what the EFT axioms provide: $Z[0]=1$, the reality condition $Z[B_{a}]^*=Z[-B_{a}]$, and $\operatorname{Im}S\ge0$ near real sources. The second, which is the subject of the paper, is \emph{Bochner positivity} or positive stochastic realizability. In this case, $Z[B_{a}]$ is the Fourier transform of a genuine probability measure, which happens if and only if
\begin{equation}
\sum_{m,n=1}^{N} \alpha_m^*\alpha_n\,Z\big[B_{a}^{(n)}-B_{a}^{(m)}\big]\;\ge\;0
\label{eq:bochner}
\end{equation}
for every finite family of real sources $B_{a}^{(n)}$ and complex coefficients $\alpha_n$. In field theory, positive definiteness on a test-function space, together with continuity, gives the Bochner-Minlos theorem \cite{SM-Minlos1959,SM-GelfandVilenkin1964}. The third, still stronger, notion is \emph{local continuum positivity}: compatibility with an independent-increments white-noise limit, which as we recall in Sec.~\ref{sec:cont} forces infinite divisibility.

Two caveats limit the scope of our models. First, Bochner positivity is not implied by quantum consistency. At finite $\hbar$ the influence functional produces correlators of non-commuting operators, and negative quasi-probabilities are legitimate quantum structure. Positivity is however a requirement whenever the noise sector is to serve as \emph{classical} stochastic forcing, which is the situation of interest here. Second, since a truncated SK exponent is only an asymptotic representation of the exact influence functional, our bounds are always formulated as constraints on \emph{completions}: they decide whether given truncated data can descend from any positive law, not whether the truncation itself is one. Finally, whenever we write KMS at finite force we implicitly assume the exponential moment
\begin{equation}
\int\!D\eta\;e^{-(F,\eta)}\,P_F(\eta)<\infty,
\label{eq:expmom}
\end{equation}
without which the complex shift in \eqref{eq:reducedKMS} is meaningless. Sec.~\ref{sec:tilt} quantifies exactly when this holds.

\section{KMS invariance as detailed balance}
\label{sec:kms}

We first establish the equivalence in the reduced convention \eqref{eq:reducedKMS}, which is the convention used throughout the main text. We then state its extension to a general time-reversal involution.

The equivalence quoted as Eq.~(4) of the main text is a direct consequence of Fourier uniqueness. Insert the transformed source into the definition \eqref{eq:Zdef} and substitute $\eta\to-\eta$:
\begin{equation}
Z_F[-B_{a}+i F]
=\int\!D\eta\;P_F(\eta)\,e^{-i(\eta,B_{a})}e^{-(F,\eta)}
=\int\!D\eta\;P_F(-\eta)\,e^{(F,\eta)}\,e^{i(\eta,B_{a})}.
\label{eq:insert}
\end{equation}
The right-hand side is the Fourier transform of the measure $e^{(F,\eta)}P_F(-\eta)$. Demanding that it coincide with $Z_F[B_{a}]$ for \emph{all} real sources is therefore, by uniqueness of Fourier transforms of finite measures, the same as demanding equality of the measures themselves,
\begin{equation}
P_F(\eta)=e^{(F,\eta)}\,P_F(-\eta),
\label{eq:DBs}
\end{equation}
and conversely, reinserting \eqref{eq:DBs} into \eqref{eq:insert} reproduces KMS invariance. It is crucial to note that positivity was \underline{not} used. Fourier uniqueness holds for finite complex measures, so the equivalence between KMS invariance and \eqref{eq:DBs} is valid for signed quasi-probabilities with finite exponential moments. This is the precise sense in which positivity is logically independent of KMS and must be imposed as a separate axiom, which is the theme of the paper.

The same argument extends immediately to the full hydrodynamic KMS transformation, including time reversal and tensor parities. These enter through a real linear involution $I$ on source space, $I^2=1$, with adjoint defined by $(I^*\eta,B)=(\eta,IB)$. For $\Theta B_{a}=IB_{a}+i F$ to square to the identity on real sources the force must be odd, $IF=-F$ (the anti-linearity of the microscopic KMS operation is already accounted for by the SK reality condition, and the reduced convention above is simply $I=-1$). Repeating the computation with the substitution $\xi=I^*\eta$, which has unit Jacobian since $I^2=1$ and in field theory is understood on cylindrical test functions with $I$ continuous on the source space, and using $(I^*\xi,F)=(\xi,IF)=-(\xi,F)$, one finds in the same way
\begin{equation}
Z_F[B_{a}]=Z_F[IB_{a}+i F]\ , \; \ \forall\,B_{a}
\qquad\Longleftrightarrow\qquad
P_F(\eta)=e^{(\eta,F)}\,P_F(I^*\eta).
\label{eq:genDB}
\end{equation}
All statements of the paper hold with ``even'' meaning $I^*$-invariant; the main text is the case $I=-1$.

Two physical situations fall outside these hypotheses and deserve mention. If the background breaks time reversal (magnetic fields, rotation), a single family $P_F$ cannot satisfy \eqref{eq:genDB}. Instead detailed balance then relates \emph{conjugate ensembles}, $P^{(B)}_F(\eta)=e^{(F,\eta)}P^{(-B)}_F(I^*\eta)$, in the manner of Onsager--Casimir, and while the tilt construction generalizes, the single-ensemble statements do not apply as written. And if the noise has power-law tails (stable L\'evy noise, say), the exponential moment \eqref{eq:expmom} fails for any $F\neq0$. The finitely many existing cumulants can still obey KMS-type relations order by order, but the finite-force equivalence with detailed balance has no content. The domain of validity of ``KMS $=$ detailed balance'' is exactly the domain of the half-force tilt, to which we now turn.

Finally, we record the relation to nonequilibrium statistical mechanics, where much of this structure is classical. For particle and energy currents in driven and mesoscopic systems, full counting statistics obeys the Gallavotti--Cohen and Lebowitz--Spohn symmetry of the cumulant generating function under $\lambda\to A-\lambda$, with $A$ the thermodynamic affinity \cite{SM-GallavottiCohen1995,SM-LebowitzSpohn1999}, equivalently the fluctuation-theorem relation $P(q)/P(-q)=e^{Aq}$ for the current distribution \cite{SM-AndrieuxGaspard2007,SM-TobiskaNazarov2005}. Equation \eqref{eq:DBs} is precisely this relation with $A\leftrightarrow F$, now identified for the SK noise sector. From that symmetry one derives hierarchies tying higher current cumulants to affinity derivatives of lower ones \cite{SM-SaitoUtsumi2008,SM-EspositoHarbolaMukamel2009}, of which the order-by-order tower of Sec.~\ref{sec:fdt} is the SK counterpart, the exponential reweighting of Sec.~\ref{sec:tilt} is the Esscher transform of classical probability \cite{SM-Esscher1932}. What is specific to the present work is threefold: the \emph{exact equivalence} of dynamical KMS and detailed balance in the LZDN sector, holding already at the level of quasi-probabilities; the characterization of the force-independent symmetric seed by the all-order gradient flow \eqref{Eq:ForceIndependentRecursion}; and the positivity theory of Secs.~\ref{sec:obst}--\ref{sec:cont}, which is independent of any fluctuation-theorem input. A covariant Crooks fluctuation theorem was recently used to derive SK-type constraints for relativistic hydrodynamics \cite{SM-MullinsHippertNoronha2025}, starting from a $\mathbb{Z}_2$ detailed-balance symmetry of the entropy-production distribution at the trajectory level. Restricted to the LZDN sector, Eq.~\eqref{eq:genDB} is the precise measure-level statement of that correspondence, upgraded to an equivalence. The notion of positivity invoked there is $\operatorname{Im}S\ge0$, guaranteed by causality and stability. The present work shows that this is strictly weaker than positive stochastic realizability, and supplies the positivity theory that a fluctuation-theorem argument alone does not provide.

\section{The half-force tilt and its domain}
\label{sec:tilt}

The general positive solution of \eqref{eq:genDB} can be parametrized as an exponential reweighting of an $I^*$-invariant seed. This invariance is not an additional assumption: the half-force untilting of any $P_F$ satisfying (S7) is automatically $I^*$-invariant. To see this constructively, let $Q_F$ be a positive $I^*$-invariant measure - $Q_F(I^*\eta)=Q_F(\eta)$ - with finite half-moment $\mathcal N[F]=\int D\eta\,Q_F\,e^{(\eta,F)/2}$, and set
\begin{equation}
P_F(\eta)=\frac{1}{\mathcal{N}[F]}\;e^{(\eta,F)/2}\,Q_F(\eta).
\label{eq:tiltdef}
\end{equation}
Using the evenness of the seed $Q_{F}$ and $(I^*\eta,F)=-(\eta,F)$ we have
	\begin{eqnarray}
		e^{(\eta,F)}P_F(I^*\eta)=\mathcal N[F]^{-1}e^{(\eta,F)}e^{-(\eta,F)/2}Q_F(\eta)=P_F(\eta)
	\end{eqnarray}
and thus detailed balance holds. Conversely, given \emph{any} normalized positive $P_F$ satisfying \eqref{eq:genDB} with $\int e^{-(\eta,F)/2}P_F<\infty$, we can define $Q_{F}(\eta)\propto e^{-(\eta,F)/2}P_F(\eta)$. Subsequently, detailed balance immediately gives $e^{-(I^*\eta,F)/2}P_F(I^*\eta)=e^{-(\eta,F)/2}P_F(\eta)$. Thus $Q_{F}$ is even, and $P_F$ is its half-force tilt.

The tilt exists at a given force $F$ whenever $\mathcal{C}_{F}[F/2]<\infty$, where
\begin{equation}
\mathcal{C}_{F}[J]\equiv \log \int D\eta\,
Q_F(\eta)e^{(J,\eta)}
\end{equation}
is the cumulant generating functional of the symmetric seed at that force. In this notation the normalization in Eq. \eqref{eq:tiltdef} is
\begin{displaymath}
\mathcal{N}[F]=e^{\mathcal{C}_{F}[F/2]} \; . 
\end{displaymath}
For a genuinely $F$-dependent family $Q_F$, the set of admissible forces need not itself be convex or origin symmetric. When the seed is force independent, $Q_F=Q_0$, the moment-generating domain of $\mathcal{C}_{0}[J]$ is convex and symmetric, and the allowed force domain is
\begin{equation}
\mathcal{D}_0=\left\{F:\mathcal{C}_{0}[F/2]<\infty\right\}.
\end{equation}
Table~\ref{tab:domain} lists representative force-independent examples.

\begin{table}[h]
\centering
\caption{Domain of the half-force tilt for representative force-independent seeds.}
\label{tab:domain}
\begin{tabular}{ll}
\toprule
Seed & Domain \\
\midrule
Gaussian & all real $F$ \\
Compact support (Bernoulli, uniform) & all real $F$ (entire MGF) \\
Compound Poisson with compact jumps & all real $F$ \\
Laplace, scale $a$ & $|F|<2/a$ \\
Power-law / stable L\'evy tails & typically no nonzero tilt \\
General L\'evy seed & $\mathcal{C}_{0}[F/2]<\infty$ \\
\bottomrule
\end{tabular}
\end{table}

\section{Master action and the gradient FDT hierarchy}
\label{sec:fdt}

For the tilted measure \eqref{eq:tiltdef}, shifting the integration exponent gives the cumulant generating functional in closed form,
\begin{equation}
\mathcal{K}_F[J]\equiv\log\!\int\!D\eta\;P_F(\eta)\,e^{(J,\eta)}
=\mathcal{C}_F\!\big[J+\tfrac F2\big]-\mathcal{C}_F\!\big[\tfrac F2\big],
\label{eq:CF}
\end{equation}
and setting $J=iB_{a}$ yields the master action, Eq.~(5) of the main text. KMS invariance is manifest: at fixed $F$, the symmetry of the seed implies $\mathcal{C}_F[-X]=\mathcal{C}_F[X]$, and therefore 
	\begin{displaymath}
		\mathcal{K}_{F}[-J-F] = \mathcal{K}_{F}[J] \; . 
	\end{displaymath}
Under \eqref{eq:reducedKMS}, $J = i B_{a}$ maps precisely to $-J-F$. Furthermore, we define the cumulants
	\begin{eqnarray}
		\kappa_{A_{1} \ldots A_{n}}[F] = \left. \partial_{J^{A_{1}}} \ldots \partial_{J^{A_{n}}} \mathcal{K}_{F}[J] \right|_{J=0} \; . 
	\end{eqnarray}
Differentiating these physical cumulants with respect to the force gives
\begin{equation}
\label{Eq:GenericCumulantRecursion}
\kappa_{A_1\cdots A_n B}(F)
=
2\frac{\partial}{\partial F^B}
\kappa_{A_1\cdots A_n}(F)
-
2\left[
\left(\frac{\partial}{\partial F^B}\right)_{J}
\frac{\partial^n \mathcal{C}_{F}[J]}
{\partial J^{A_1}\cdots\partial J^{A_n}}
\right]_{J=F/2},
\end{equation}
where $\left(\partial/\partial F^B\right)_J$ acts only on the explicit
$F$-dependence of the seed $Q_F$, with its argument $J$ held fixed.
This is the tensorial form of Eq.~(6) of the main text. For a
force-independent seed, $Q_F=Q_0$, equivalently $\mathcal{C}_{F}=\mathcal{C}_{0}$, the second
term in Eq.~\eqref{Eq:GenericCumulantRecursion} vanishes and the hierarchy reduces to
\begin{equation}
\label{Eq:ForceIndependentRecursion}
\frac{\partial}{\partial F^B}
\kappa_{A_1\cdots A_n}(F)
=
\frac{1}{2}\kappa_{A_1\cdots A_n B}(F),
\qquad n\geq 1\; .
\end{equation}
Thus a force-independent seed implies the gradient flow.

The converse is also useful: the gradient flow characterizes a
force-independent seed among detailed-balanced families. Suppose that a
smooth detailed-balanced family $\{P_F\}$ obeys Eq.~\eqref{Eq:ForceIndependentRecursion}, and that
$\mathcal{K}_{F}[J]$ is analytic in $J$ in a common neighbourhood of $J=0$ and
sufficiently smooth in $F$ to interchange the required derivatives.
Equation~\eqref{Eq:ForceIndependentRecursion} then implies equality of all $J$-Taylor coefficients of
the functional relation
\begin{equation}
\label{Eq:MasterRecursion}
2\frac{\partial \mathcal{K}_{F}[J]}{\partial F^B}
=
\frac{\partial \mathcal{K}_{F}[J]}{\partial J^B}
-
\kappa_B(F),
\end{equation}
including the zeroth-order coefficient, since $\mathcal{K}_{F}[0]=0$. Analyticity in
$J$ therefore promotes the cumulant hierarchy to Eq.~\eqref{Eq:MasterRecursion} in that
neighbourhood.

To solve Eq.~\eqref{Eq:MasterRecursion}, define
\[
G[u,F]\equiv \mathcal{K}_{F}[u-F/2].
\]
Using Eq.~\eqref{Eq:MasterRecursion} gives
\[
\frac{\partial G[u,F]}{\partial F^B}
=
-\frac{1}{2}\kappa_B(F),
\]
which is independent of $u$. Hence
\[
G[u,F]=\mathcal{C}_{0}[u]-c(F)
\]
for some $F$-independent functional $\mathcal{C}_{0}[u]$. The normalization
$\mathcal{K}_{F}[0]=0$ fixes $c(F)=\mathcal{C}_{0}[F/2]$, and therefore
\[
\mathcal{K}_{F}[J]
=
\mathcal{C}_{0}\!\left[J+\frac{F}{2}\right]
-
\mathcal{C}_{0}\!\left[\frac{F}{2}\right].
\]
Detailed balance, equivalently $\mathcal{K}_{F}[-J-F]=\mathcal{K}_{F}[J]$, then implies
$\mathcal{C}_{0}[-J]=\mathcal{C}_{0}[J]$. Thus $\mathcal{C}_{0}$ is the cumulant generating functional of
an even, force-independent seed $Q_0$, with
\[
Q_0(\eta)\propto e^{-(F,\eta)/2}P_F(\eta).
\]
Consequently, within the stated regularity assumptions, Eq.~\eqref{Eq:ForceIndependentRecursion} is
both necessary and sufficient for a detailed-balanced family to arise
from a force-independent symmetric seed.

The moral spelled out in the main text follows: the gradient flow is \emph{not} a consequence of KMS alone. What KMS by itself gives, in one dimension, is obtained by expanding
$\mathcal{K}_{F}(J)=\mathcal{K}_{F}(-J-F)$ in powers of $J$:
\begin{equation}
\kappa_m(F)=(-1)^m\sum_{r\ge0}\frac{(-1)^r}{r!}\,\kappa_{m+r}(F)\,F^r,
\label{eq:tower}
\end{equation}
whose first entries, $2\kappa_1=\kappa_2F-\tfrac12\kappa_3F^2+\tfrac16\kappa_4F^3-\dots$ and $0=-\kappa_3F+\tfrac12\kappa_4F^2-\dots$, are useful diagnostics for any proposed non-Gaussian SK action but are strictly weaker than \eqref{Eq:ForceIndependentRecursion}. A violation of \eqref{Eq:ForceIndependentRecursion} at some order in $F$ therefore signals a state-dressed ($F$-dependent) seed, not a KMS violation. For a force-independent seed with zero-force cumulants $\kappa^{(n)}$ (odd ones vanishing), resumming \eqref{Eq:ForceIndependentRecursion} gives the exact expansion
$\kappa_{A_1\ldots A_n}(F)=\sum_{r\ge0}\tfrac{1}{2^r r!}\,\kappa^{(n+r)}_{A_1\ldots A_nB_1\ldots B_r}F^{B_1}\!\ldots F^{B_r}$,
whose lowest entries are the linear FDT $\kappa_A=\tfrac12 \kappa^{(2)}_{AB}F^B+\tfrac1{48} \kappa^{(4)}_{ABCD}F^BF^CF^D+\dots$ and the exact relation $\kappa_{ABC}(F)=2\,\partial\kappa_{AB}/\partial F^C$ quoted in the main text. Note the rigidity: once $\kappa^{(4)}\neq0$ is chosen, the cubic drift $\tfrac1{48} \kappa^{(4)}F^3$ is not optional.

\section{What positivity forbids}
\label{sec:obst}

\emph{No exact polynomial non-Gaussianity.} Suppose $Z[B_{a}]$ comes from a positive measure, and along some real source direction $\varphi$ the restriction $Z(t)=Z[t\varphi]$ has the form $e^{P(t)}$ with $P$ a polynomial. The pushforward of the measure along $\eta\mapsto(\eta,\varphi)$ is then a one-dimensional probability law whose characteristic function is $e^{P(t)}$, and Marcinkiewicz's theorem \cite{SM-Lukacs1970,SM-RajagopalSudarshan1974} states that this is possible only for $\deg P\le2$. In particular the KMS-invariant functional $\exp[-\tfrac N2T-\lambda T^2]$, with $T(b,F)=b(b-i F)$, restricts at $F=0$ to $\exp(-\tfrac N2b^2-\lambda b^4)$ and is therefore not positive for any $\lambda\neq0$, which is the negativity displayed in Fig.~1 of the main text and the reason all finite polynomial vertices must be read as truncations. (A related but logically distinct fact for generator truncations is the Pawula theorem \cite{SM-Pawula1967}.)

\emph{The fourth-cumulant bound.} At finite force the noise need not be symmetric, so let
\[
Y \equiv X-\langle X\rangle
\]
be the centered random variable. Its first four central moments are
\[
\mu_2=\kappa_2,\qquad
\mu_3=\kappa_3,\qquad
\mu_4=\kappa_4+3\kappa_2^2.
\]
Positivity of the probability measure implies positivity of the moment
matrix associated with the polynomials $1,Y,Y^2$,
\[
M=
\begin{pmatrix}
1 & 0 & \mu_2\\
0 & \mu_2 & \mu_3\\
\mu_2 & \mu_3 & \mu_4
\end{pmatrix}
\succeq 0 .
\]
Hence $\det M\geq0$, which gives
\[
\mu_2\mu_4-\mu_3^2-\mu_2^3\geq0.
\]
Substituting the central moments in terms of cumulants yields
\begin{equation}
\kappa_2(F)\kappa_4(F)
+2\kappa_2(F)^3
-\kappa_3(F)^2
\geq0,
\end{equation}
or, for $\kappa_2(F)>0$,
\begin{equation}
\kappa_4(F)\geq
\frac{\kappa_3(F)^2}{\kappa_2(F)}
-2\kappa_2(F)^2,
\end{equation}
which is Eq.~(7) of the main text. The degenerate case
$\kappa_2=0$ corresponds to a deterministic variable and is trivial.

At zero force detailed balance makes the distribution even, so
$\kappa_3(0)=0$, and the bound reduces to
\begin{equation}
\kappa_4\geq -2\kappa_2^2.
\label{eq:k4S} 
\end{equation}
Equality at zero force requires $Y^2$ to be almost surely constant,
and hence is attained by the symmetric two-point law $Y=\pm q$ with
equal probability. For the quartic truncation,
$\kappa_2=N$ and $\kappa_4=-24\lambda$, giving
$\lambda\leq N^2/12$ as quoted in the main text. Applied to
$X=v_A\eta^A$, the same bound holds directionally for
multi-component noise.

\emph{The sixth-cumulant bound.} One level up, apply the Cauchy--Schwarz inequality to the pair $(X,X^3)$: $m_4^2=\mathbb E[X \cdot X^3]^2\le m_2m_6$. Substituting the even-variable conversions $m_4=\kappa_4+3\kappa_2^2$ and $m_6=\kappa_6+15\kappa_2\kappa_4+15\kappa_2^3$ gives
\begin{equation}
0\;\le\;m_2m_6-m_4^2\;=\;\kappa_2\kappa_6+9\kappa_2^2\kappa_4+6\kappa_2^4-\kappa_4^2,
\label{eq:k6S}
\end{equation}
which is Eq.~(8) of the main text; equality requires $X^3\propto X$, i.e.\ support on $\{0,\pm q\}$. In particular, let
	\begin{equation}
P(X=0)=1-p,
\qquad
P(X=\pm q)=\frac{p}{2},
\end{equation}
and then
\begin{equation}
q^2=
\frac{\kappa_4+3\kappa_2^2}{\kappa_2},
\qquad
p=
\frac{\kappa_2^2}
{\kappa_4+3\kappa_2^2} \; .
\end{equation}
This saturates the bound \eqref{eq:k6S}. In particular, the Bernoulli law
\begin{equation}
p=1
\qquad\Longleftrightarrow\qquad
\kappa_4=-2\kappa_2^2,
\end{equation}
is a special sub-case of this saturation.

In summary: in the force-free case, a prescribed negative $\kappa_4$ forces $\kappa_6\ge(\kappa_4^2-9\kappa_2^2\kappa_4-6\kappa_2^4)/\kappa_2$, strictly positive once $\kappa_4$ is negative enough: positivity therefore correlates the sixth-order vertex with the lower
cumulants, although it does not determine it uniquely. For the two-point law all moments are $m_{2k}=q^{2k}$, so \emph{both} left-hand sides above vanish identically: the Bernoulli law saturates the two bounds simultaneously and is the extreme point of the completion region, which is why it reappears as the matching example throughout. Higher orders are policed by positive semi-definiteness of the Hankel moment matrices and their shifted companions, which give necessary conditions on $\kappa_8,\kappa_{10},\dots$ for any displayed truncation. One caveat: the bounds of this section refer to \emph{cell} variables and hence constrain the pair (truncated data, regularization cell); under white-noise scaling their content depends on the cell volume, and only the continuum densities of the next section are scheme independent.

\section{The continuum limit: Hamburger and Stieltjes hierarchies}
\label{sec:cont}

In field theory, a functional on a test-function space that is normalized, positive definite in the sense \eqref{eq:bochner}, and continuous is the characteristic functional of a probability measure on the dual space, by the Bochner--Minlos theorem \cite{SM-Minlos1959,SM-GelfandVilenkin1964}. Local white noise is a stronger notion: independence of the noise on disjoint regions makes the cumulant functional additive over disjoint supports, and under standard regularity assumptions \cite{SM-RajputRosinski1989,SM-DalangHumeau2017} this forces a \emph{local} exponent, $\mathcal{K}_{F}[J]=\int\mathrm{d}^{d+1}x\;\psi_{x,F}(J(x))$, with each $\psi_{x,F}$ of L\'evy--Khintchine (infinitely divisible) form. In the symmetric class with exponential moments,
\begin{equation}
\psi(J)=\tfrac12\,J_AG^{AB}J_B+\int\nu(\mathrm{d} q)\,\big[\cosh(q^AJ_A)-1\big]:
\label{eq:psi}
\end{equation}
a Gaussian part of covariance density $G$, plus independently
scattered jumps governed by the L\'evy measure $\nu(\mathrm{d} q)$. More generally, at fixed force $F$, let $G(F)$ and $\nu_F$ denote the Gaussian variance density and Lévy measure of the physical infinitely divisible law $P_F$. In the
scalar homogeneous case, cumulants of the noise integrated over a cell
of volume $V_{\mathrm{cell}}$ scale as
\[
\kappa_{k}(Y_\Omega;F)=c_{k}(F) V_{\mathrm{cell}} +o(V_{\mathrm{cell}}),
\]
with
\begin{equation}
c_{k}(F) = G(F) \delta_{k,2} + \int_{\mathbbm{R}} q^{k} \nu_{F}(\mathrm{d} q) \; , \qquad k \geq 2 \; . 
\label{eq:cumulantdensitiesS}
\end{equation}
Thus the Gaussian component contributes only to the second cumulant,
whereas all cumulants of order $k\geq 3$ come entirely from the jump measure.

This scaling turns the finite-cutoff bounds into continuum statements
within the independently scattered white-noise class. Inserting it
into Eq.~\eqref{eq:k4S}, dividing by $V_{\mathrm{cell}}$, and taking $V_{\mathrm{cell}}\to0$
gives $c_4\geq0$. Thus a negative fourth-cumulant density is excluded
under the assumptions of independent increments and linear volume
scaling, but not for more general colored or correlated continuum
noises. The same insertion into Eq.~\eqref{eq:k6S} gives, after
dividing by $V_{\mathrm{cell}}^2$, the density bound $c_2c_6\geq c_4^2$.

These are the first two entries of an exact moment structure. To treat
the Gaussian and jump contributions uniformly, define a positive
measure $\widetilde{\sigma}_{F}$ on $\mathbbm{R}$ by
\begin{equation}
\widetilde{\sigma}_{F}(\mathrm{d} q)
=
G(F) \delta_{0}(\mathrm{d} q)
+
q^2 \,\nu_{F}(\mathrm{d} q) \; . 
\label{eq:sigmadefS}
\end{equation}
Equation
\eqref{eq:cumulantdensitiesS} then gives
\begin{equation}
c_{k+2}(F)
=
\int_{-\infty}^{\infty} q^k\,\widetilde{\sigma}_{F}(\mathrm{d} q),
\qquad
k=0,1,2,\ldots .
\label{eq:stieltjesS}
\end{equation}
Indeed, for $k=0$ the right-hand side is
\[
G(F) +\int q^2\,\nu_{F}(\mathrm{d} q)=c_2(F),
\]
whereas for $k\geq1$ the term $G$ does not contribute and
\[
\int_{-\infty}^{\infty} q^k\,\widetilde{\sigma}_{F}(\mathrm{d} q)
=
\int q^{k+2} \,\nu_{F}(\mathrm{d} q)
=
c_{k+2}(F).
\]

The set $\{c_{k+2}(F)\}_{k\geq0}$ is a Hamburger moment sequence. Consequently, the
Hankel matrix satisfies
\[
\begin{pmatrix}
c_2(F) & c_3(F)\\
c_3(F) & c_4(F)
\end{pmatrix}
\succeq 0,
\]
and therefore
\begin{equation}
c_2(F)c_4(F)\geq c_3(F)^2.
\end{equation}
For a force-independent seed, $c_3(F)=2\,\partial_F c_2(F)$, giving
\begin{equation}
c_4(F)\geq
\frac{4[\partial_F c_2(F)]^2}{c_2(F)} 
\end{equation}
as displayed in the main text.

At zero force, detailed balance implies that the L\'evy measure is
symmetric, $\nu_0(dq)=\nu_0(-dq)$, and hence all odd cumulant densities
vanish. The even subsequence admits a stronger representation on the
positive half-line. Define a positive measure $\rho$ on $[0,\infty)$ by
\begin{equation}
\int_0^\infty f(s)\,\rho(ds)
=
G f(0)
+
\int_{\mathbb R} q^2 f(q^2)\,\nu_0(dq)
\end{equation}
for every suitable test function $f$. Thus $\rho$ consists of a term $G$ at the origin together with the pushforward of
$q^2\nu_0(dq)$ under $q\mapsto q^2$. It follows that
\begin{equation}
c_{2k+2}(0)
=
\int_0^\infty s^k\,\rho(ds),
\qquad
k=0,1,2,\ldots .
\end{equation}
Indeed, for $k=0$ this gives
\[
c_2(0)=G +\int_{\mathbb R}q^2\,\nu_0(dq),
\]
while for $k\geq1$ the term at the origin does not contribute and
\[
\int_0^\infty s^k\,\rho(ds)
=
\int_{\mathbb R}q^{2k+2}\,\nu_0(dq)
=
c_{2k+2}(0).
\]
Consequently,
\[
\bigl(c_2(0),c_4(0),c_6(0),\ldots\bigr)
\]
is a Stieltjes moment sequence. The classical solvability criterion for
the infinite Stieltjes moment problem~\cite{SM-Akhiezer1965} is positive
semidefiniteness of the two families of Hankel matrices
\begin{equation}
\label{Eq:HankelConditions}
H^{(0)}_{ij}=c_{2(i+j)+2}(0),
\qquad
H^{(1)}_{ij}=c_{2(i+j)+4}(0),
\qquad i,j\geq0 .
\end{equation}
Their first nontrivial principal minors give
\[
c_2c_6\geq c_4^2,
\qquad
c_4c_8\geq c_6^2,
\qquad
c_6c_{10}\geq c_8^2,
\]
with all cumulant densities in this line evaluated at $F=0$.

Conversely, any complete sequence satisfying the Stieltjes criterion
admits a representing positive measure $\rho$ on $[0,\infty)$. Its
mass at the origin defines the Gaussian variance density,
\[
G =\rho(\{0\}),
\]
while its restriction to $(0,\infty)$ reconstructs a symmetric jump
measure by undoing the $q^2$ weighting and assigning equal weight to
$q=\pm\sqrt{s}$. The resulting Gaussian-plus-L\'evy white noise
therefore realizes the prescribed zero-force even cumulant densities.
For a finite list of cumulants, the same conditions are truncated
moment constraints, with sufficiency understood as the existence of a
positive Stieltjes extension.

For a complete scalar sequence at zero force, the Stieltjes theorem makes the two Hankel
conditions \eqref{Eq:HankelConditions} both necessary and sufficient for the existence of a positive measure $\rho$,
and hence, through the construction above, for positive independent-increment realizability.
For a finite list of cumulants, they are the corresponding truncated
moment constraints, with sufficiency understood as the existence of a
positive Stieltjes extension. The representing measure need not be
unique when the moment problem is indeterminate. In the
multi-component case, directional Hankel conditions remain necessary
but are not sufficient in general.

The hypotheses should be kept in view: independence across cells, and cumulants linear in cell volume. Both fail in critical regimes, where fluctuations are correlated over a length $\xi$ and cumulants scale with the correlation volume; the constraints of this section delimit the strictly ultralocal limit, as emphasized in the main text.

\section{A library of exact seeds}
\label{sec:lib}

Each force-independent seed $Q_0$ below with $\mathcal{C}_{F}=\mathcal{C}_{0}$, inserted in \eqref{eq:CF}, yields an exact, closed-form, KMS-invariant, positively realizable SK action. All are one-dimensional (per component or per jump direction); coordinates $(x,y)=(\kappa_4/\kappa_2^2,\kappa_6/\kappa_2^3)$ refer to Fig.~2 of the main text.

\emph{Tilted Laplace.} The two-sided exponential $P_0=\tfrac1{2a}e^{-|\eta|/a}$ has $\mathcal{C}_0(J)=-\log(1-a^2J^2)$ and tilt domain $|F|<2/a$. All cumulants are elementary,
\begin{equation}
\kappa_n(F)=(n-1)!\;a^n\left[\frac{1}{(1-aF/2)^n}+\frac{(-1)^n}{(1+aF/2)^n}\right],
\label{eq:laplace}
\end{equation}
and one checks $2\kappa_n'(F)=\kappa_{n+1}(F)$ directly, a useful sanity test of the whole formalism. Values at $F=0$, $\kappa_{2n}(0)=2(2n-1)!\,a^{2n}$ put it at $(x,y)=(3,30)$.

\emph{Bernoulli.} The two-point law $P_0=\tfrac12\delta(\eta-q)+\tfrac12\delta(\eta+q)$ has the entire exponent $\mathcal{C}_0(J)=\log\cosh(qJ)$. With an independent Gaussian residual of variance $G\ge0$ the $F=0$ data are $\kappa_2(0)=G+q^2$ and $\kappa_4(0)=-2q^4<0$, so a negative quartic truncation $(N,\lambda)$ is matched \emph{exactly through fourth order} by $q^4=12\lambda$ and $G=N-\sqrt{12\lambda}$, which is non-negative precisely when $\lambda\le N^2/12$: the positivity condition of Sec.~\ref{sec:obst} is satisfied, and it
is saturated only at $G=0$, equivalently
$\lambda=N^2/12$. The pure two-point seed $G=0$ also saturates the sixth-cumulant
bound. The two-point law is the extremal completion saturating the
fourth-cumulant bound; away from saturation, admissible
negative-$\kappa_4$ data generally admit nonunique positive
completions (see the dashed curve of Fig. 1 in the main text). One structural point: at $F=0$ the characteristic function $e^{-Gb^2/2}\cos(qb)$ has zeros, so the \emph{action} is only locally defined; relatedly, two-point and uniform seeds are not infinitely divisible and hence admit no continuum limit. They are finite-cutoff moment-matching tools, the honest probabilistic counterpart of a negative-$\kappa_4$ truncation, while the continuum-compatible members of the library are the Gaussian-plus-L\'evy classes.

\emph{Uniform.} $P_0=\tfrac1{2q}\mathbf 1_{[-q,q]}$, $\mathcal{C}_0(J)=\log[\sinh(qJ)/(qJ)]$: a smooth finite-cutoff completion of negative $\kappa_4$, with $\kappa_2(0)=q^2/3$, $\kappa_4(0)=-2q^4/15$, $\kappa_6(0)=16q^6/63$, i.e.\ $(x,y)=(-6/5,48/7)$, and the correspondingly stricter matching bound $\lambda\le N^2/20$.

\emph{Symmetric compound Poisson (Skellam).} Kicks $\pm q$ at equal rate $r/2$ give $\mathcal{C}_0(J)=r[\cosh(qJ)-1]$ and $F=0$ cumulants $\kappa_2(0)=G+rq^2$, $\kappa_{2n}(0)=rq^{2n}$ for $n\ge2$; eliminating $r$ shows the pure-jump family traces exactly the parabola $y=x^2$ of Fig.~2, the continuum boundary, corresponding to a mono-atomic intensity measure $\sigma$. The characteristic function is globally nonzero (infinite divisibility): this is the clean completion class for positive higher even cumulants and, consistently with Sec.~\ref{sec:cont}, it cannot match negative $\kappa_4$.

\section{Hydrodynamic tensor matching}
\label{sec:hydro}

\emph{Charge diffusion.} In a local rest frame the dissipative current noise has the Landau--Lifshitz kernel $N^{\rm target}_{ij}=2T\sigma\,\delta_{ij}$. Take kicks of fixed magnitude $q_0$ in uniformly random directions $n\in S^{d-1}$ at total rate $r$. Isotropy fixes the sphere moments $\langle n_in_j\rangle=\delta_{ij}/d$ and $\langle n_in_jn_kn_l\rangle=(\delta_{ij}\delta_{kl}+\delta_{ik}\delta_{jl}+\delta_{il}\delta_{jk})/[d(d+2)]$, so the kick sector carries covariance $rq_0^2\delta_{ij}/d$ and quartic vertex $c_{ijkl}(0)=\tfrac{rq_0^4}{d(d+2)}(\delta\delta+\delta\delta+\delta\delta)_{ijkl}$, and the Gaussian residual required to match the target is
\begin{equation}
G_{ij}=\Big(2T\sigma-\frac{rq_0^2}{d}\Big)\,\delta_{ij}\;\succeq\;0
\qquad\Longleftrightarrow\qquad
\frac{rq_0^2}{d}\le 2T\sigma.
\label{eq:residual}
\end{equation}
The seed exponent is exact and closed form: resumming $\langle n_1^{2m}\rangle=\Gamma(\tfrac d2)(2m)!/[4^mm!\,\Gamma(m+\tfrac d2)]$ gives
\begin{equation}
\mathcal{C}_0[J]=\tfrac12 J_iG_{ij}J_j+r\big[\Phi_d(q_0|J|)-1\big],\qquad
\Phi_d(z)=\Gamma\!\big(\tfrac d2\big)\Big(\tfrac2z\Big)^{d/2-1}I_{d/2-1}(z),
\label{eq:Phid}
\end{equation}
with $\Phi_3(z)=\sinh z/z$. The mean tilted drift, the exact nonlinear response of this completion, follows by isotropy: $c_i(F)=\hat F_i[\tfrac12(2T\sigma-\tfrac{rq_0^2}{d})|F|+rq_0\Phi_d'(\tfrac{q_0|F|}{2})]$, reducing to the FDT value $T\sigma F_i$ at small force and growing exponentially once $q_0|F|\gtrsim1$. As stressed in the main text, this far-from-equilibrium profile is a property of the chosen completion, while the gradient relations of Sec.~\ref{sec:fdt} are shared by every fixed-seed positive noise.

\emph{Stress noise.} Decompose symmetric spatial tensors with the shear and trace projectors $P^{\rm ST}_{ij,kl}=\tfrac12(\delta_{ik}\delta_{jl}+\delta_{il}\delta_{jk})-\tfrac1d\delta_{ij}\delta_{kl}$ and $P^{\rm tr}_{ij,kl}=\tfrac1d\delta_{ij}\delta_{kl}$, with $\operatorname{tr}P^{\rm ST}=D_{\rm ST}=d(d+1)/2-1$; the Landau--Lifshitz target is $N^{\rm target}=4T\eta\,P^{\rm ST}+2Td\zeta\,P^{\rm tr}$. A convenient positive seed is a sum of independent sectors: bulk kicks $q^{\rm bulk}_{ij}=s\,\delta_{ij}/\sqrt d$ at rate $r_b$, contributing $r_bs^2P^{\rm tr}$, and shear kicks $q^{\rm shear}_{ij}=q_s\,n_{ij}$ at rate $r_s$, with $n$ uniform on the unit sphere of the $D_{\rm ST}$-dimensional shear space, contributing $\tfrac{r_sq_s^2}{D_{\rm ST}}P^{\rm ST}$. Because $P^{\rm ST}$ is the metric of that Euclidean space, the same isotropy argument as above gives the fourth shear cumulant as the $P^{\rm ST}$-triplet analogue of the vector formula, and the residual condition
\begin{equation}
G=\Big(4T\eta-\frac{r_sq_s^2}{D_{\rm ST}}\Big)P^{\rm ST}
+\big(2Td\zeta-r_bs^2\big)P^{\rm tr}\;\succeq\;0.
\label{eq:stressres}
\end{equation}
The shear-sector generating function is $\Phi_{D_{\rm ST}}$ of \eqref{eq:Phid}, and the bulk sector is a one-dimensional Skellam exponent.

\section{Tilted-kick sampling and Monte Carlo details}
\label{sec:mc}

The practical value of the L\'evy class rests on a closure property, quoted as Eq.~(14) of the main text: the half-force tilt of a L\'evy seed is again a L\'evy noise. This is a two-line computation from \eqref{eq:CF}. For $\mathcal{C}_0[J]=\tfrac12 J_AG^{AB}J_B+\int\nu(\mathrm{d} q)[e^{(q,J)}-1]$ with symmetric $\nu$,
\begin{equation}
\mathcal{K}_F[J]=\tfrac12\,JGJ+\tfrac12\,(GF,J)
+\int \nu_F(\mathrm{d} q)\,\big[e^{(q,J)}-1\big],
\qquad
\nu_F(\mathrm{d} q)=e^{(q,F)/2}\,\nu(\mathrm{d} q),
\label{eq:tiltlevy}
\end{equation}
which is again of L\'evy--Khintchine form: an independent Gaussian with mean $\tfrac12GF$ and covariance $G$, plus kicks with exponentially reweighted rates. Detailed balance is thus implemented microscopically in the most physical way possible, since the force does not deform the kicks but reweights their rates, and \eqref{eq:tiltlevy} \emph{is} the sampling algorithm. For the isotropic $d=3$ seed the tilted total rate is $r_F=r\,\Phi_3(z)$ with $z=q_0F/2$, and the kick direction has density $\propto e^{zu}$ in $u=\cos\theta\in[-1,1]$, sampled by the explicit inverse CDF $u=z^{-1}\log[e^{-z}+\xi(e^z-e^{-z})]$ with $\xi$ uniform on $[0,1]$. Per cell of volume $V_{\mathrm{cell}}$: draw the Gaussian part from $\mathcal N(\tfrac12GF V_{\mathrm{cell}},\,G V_{\mathrm{cell}})$; draw the kick number $K$ from $\mathrm{Poisson}(r_F V_{\mathrm{cell}})$; draw $K$ directions from the tilted density and add $q_0n^{(i)}$. No reweighting, no negative weights. The two ingredients used here, the closure of the L\'evy class under exponential tilting and the shot-noise representation for sampling infinitely divisible laws, are classical in probability \cite{SM-Sato1999,SM-RajputRosinski1989}; what is specific to the present setting is their combination with dynamical KMS, which fixes the tilt to be by half the force and thereby ties the sampling weights to detailed balance.

Along the force the exact tilted cumulant densities of this completion are
\begin{equation}
c_1(F)=\frac{1}{2}GF+r q_0\Phi_3'(z),\qquad
c_k(F)=G\delta_{k,2}+r q_0^k\Phi_3^{(k)}(z)
\quad (k\geq2),
\label{eq:workedS}
\end{equation}
which satisfy $2\partial_F c_k=c_{k+1}$ identically in $F$.
For a coarse-graining cell of spacetime volume $V_{\mathrm{cell}}$,
the corresponding cell cumulants are
\[
\kappa_k(F)=V_{\mathrm{cell}}\,c_k(F).
\]
The Monte Carlo of the main text uses $V_{\mathrm{cell}}=1$, so the
numerical values of $c_k$ and $\kappa_k$ coincide there. At the parameters of the main text ($T=1$, $q_0=1$, $r=3/2$, $G=1/2$, so half the zero-force variance is carried by kicks) and at $F=1$ (i.e.\ $z=\tfrac12$) they evaluate to the elementary closed forms
$c_1=\tfrac14+3\cosh\tfrac12-6\sinh\tfrac12$,
$c_2=\tfrac12+27\sinh\tfrac12-12\cosh\tfrac12$,
$c_3=75\cosh\tfrac12-162\sinh\tfrac12$,
$c_4=1299\sinh\tfrac12-600\cosh\tfrac12$,
numerically $(0.506306,\,1.038062,\,0.154508,\,0.327223)$. The $F=0$ densities are $c_2=G+rq_0^2/3=1$, $c_4=rq_0^4/5=3/10$, $c_6=rq_0^6/7=3/14$, and the first Stieltjes minors of Sec.~\ref{sec:cont} are
\[
c_2c_6-c_4^2=\frac{87}{700}>0,
\qquad
c_4c_8-c_6^2=\frac{1}{245}>0.
\]
They are strictly positive because the representing measure contains
both a Gaussian term at $s=0$ and a nontrivial distribution of jump
intensities.

The Monte Carlo of the main text samples $2\times10^7$ cells at $F=1$, split into $20$ independent batches. Within each batch the cumulants are estimated from central moments ($\kappa_4=m_4^c-3(m_2^c)^2$, and so on); the quoted central value is the mean over batches, and the quoted uncertainty is the sample standard deviation across batches (with Bessel's correction) divided by $\sqrt{20}$. As an independent cross-check of the hierarchy on sampled data alone,
we have also measured $\kappa_2(F)$ on the grid
$F\in\{0.8,1.2,1.6\}$ and compared twice its central finite difference
at $F=1.2$ with the third cumulant $\kappa_3(1.2)$ measured at the same
force. The two agree within one standard deviation (pull $-0.1$).
The centered finite difference carries the expected $O(h^2)$
discretization bias, so this comparison should be regarded as an
independent consistency check rather than a precision test of the
hierarchy. No exact input enters either side of the comparison.

\section{Confronting the microscopic vertices}
\label{sec:micro}

Here we work out the two comparisons summarized in the main text. Both use only the low-order noise data of the cited references and the bounds of Secs.~\ref{sec:obst} and \ref{sec:cont}.

\emph{Brownian particle of Lin, Bu, and Lei.} The effective Lagrangian of \cite{SM-LinBuLei2024} reads $L=i T\Delta_a^2-\Delta_a\partial_t\Delta_r-m\Delta_a\Delta_r+i\epsilon_1\Delta_a^4-\epsilon_2\Delta_a^3\Delta_r+\dots$, with $\Delta_a$ the difference source. Its pure-noise part, obtained at $\Delta_r=0$, is $i S=-T\Delta_a^2-\epsilon_1\Delta_a^4$ per unit time, so in the notation of the main text the noise strength and quartic coupling are
\begin{equation}
N=2T,\qquad \lambda=\epsilon_1 ,
\end{equation}
as densities; over a coarse-graining step $\delta t$ the single-cell cumulants are $\kappa_2=2T\delta t$ and $\kappa_4=-24\epsilon_1\delta t$. Stability of their Fokker--Planck equation requires $\epsilon_1>0$, hence $\kappa_4<0$. Reference \cite{SM-LinBuLei2024} isolates the quartic vertex as a separate noise $\eta$ with $\langle\eta^2\rangle=0$ and $\langle\eta^4\rangle=-24\epsilon_1/\delta t^3$, notes that its weight function is not positive, and observes that this is unavoidable. Our Eq.~\eqref{eq:k4S} explains and sharpens the point. With $\kappa_2=0$ the bound $\kappa_4\ge-2\kappa_2^2$ forces $\kappa_4\ge0$, so any splitting that isolates the non-Gaussian vertex from the variance is necessarily non-positive; the meaningful object is the \emph{total} single-step noise, of variance $\kappa_2=2T\delta t$. Positivity of the total noise through fourth order, $\kappa_4\ge-2\kappa_2^2$, reads $-24\epsilon_1\delta t\ge-2(2T\delta t)^2$, that is
\begin{equation}
\boxed{\;\delta t\;\ge\;\frac{3\epsilon_1}{T^2}\;}
\label{eq:dtboundS}
\end{equation}
a sharp lower bound on the coarse-graining time, saturated by the Bernoulli completion of Sec.~\ref{sec:lib} with kick amplitude $q=(12\epsilon_1\delta t)^{1/4}$ and Gaussian residual $G=2T\delta t-\sqrt{12\epsilon_1\delta t}$. In the window
\begin{equation}
\frac{3\epsilon_1}{T^2}\;\le\;\delta t\;<\;\frac{9+\sqrt{105}}{2}\,\frac{\epsilon_1}{T^2}\approx 9.62\,\frac{\epsilon_1}{T^2},
\end{equation}
the normalized sixth cumulant $y=\kappa_6/\kappa_2^3$ of any positive completion must be strictly positive, by \eqref{eq:k6S} evaluated at $x=\kappa_4/\kappa_2^2=-6\epsilon_1/(T^2\delta t)$; a truncation with $\kappa_6=0$ is then not completable through sixth order there. Finally, the ambiguity that \cite{SM-LinBuLei2024} reports as $\delta t$ and $\epsilon_1$ tend to zero together, and the divergences of continuum perturbative treatments they discuss, are structural from the present viewpoint: the fourth-cumulant density is $c_4=-24\epsilon_1<0$, so by the continuum argument of Sec.~\ref{sec:cont} no positive independent-increments continuum limit exists. For $\delta t$ satisfying \eqref{eq:dtboundS}, the Bernoulli-plus-Gaussian finite-cell
completion of Sec.~\ref{sec:lib} provides a directly sampleable positive
realization of the combined noise.

\emph{Holographic SU(2) diffusion of Bu, Sun, and Zhang.} Reference \cite{SM-BuSunZhang2022} computes, from a probe SU(2) gauge field in doubled Schwarzschild-AdS$_5$, the quartic terms of the diffusion effective action analytically. The purely spatial noise couplings, quartic in the difference sources $B^{\rm a}_{ai}$ (with $\rm a$ the flavor index), appear there as $\chi_{3,4}(B^{\rm a}_{ai}B^{\rm a}_{ai})^2+\chi_{4,4}B^{\rm a}_{ai}B^{\rm b}_{ai}B^{\rm a}_{aj}B^{\rm b}_{aj}$ with $\chi_{3,4}=-\chi_{4,4}=-i\pi/128$. Introduce the positive-semidefinite flavor matrix
\begin{equation}
M^{\rm ab}=\sum_i B^{\rm a}_{ai}B^{\rm b}_{ai},
\end{equation}
so that the two structures are $(\operatorname{tr}M)^2$ and $\operatorname{tr}(M^2)$. The corresponding contribution to $i S$ is
\begin{equation}
i S\supset \frac{\pi}{128}\big[(\operatorname{tr}M)^2-\operatorname{tr}(M^2)\big].
\end{equation}
To probe the quartic form along an arbitrary direction in the combined flavor-spatial source space, set
\[
B^{\rm a}_{ai}=b\,V^{\rm a}_i,
\qquad
\sum_{{\rm a},i}(V^{\rm a}_i)^2=1,
\]
and introduce the normalized flavor Gram matrix
\[
(M_V)^{{\rm ab}}
\equiv
\sum_iV^{\rm a}_iV^{\rm b}_i.
\]
Then
\[
M=b^2M_V,
\qquad
\operatorname{tr}M_V=1,
\]
and the quartic contribution becomes
\[
i S_4
=
\frac{\pi}{128}b^4
\left[
1-\operatorname{tr}(M_V^2)
\right].
\]
Using
\[
\log Z(b)
=
\sum_{n=1}^{\infty}
\frac{(i b)^n}{n!}\kappa_n,
\]
the corresponding fourth-cumulant density is therefore
\begin{equation}
c_4(V)
=
\frac{3\pi}{16}
\left[
1-\operatorname{tr}(M_V^2)
\right].
\label{eq:holoS}
\end{equation}
The matrix $M_V$ is positive semidefinite. Denoting its eigenvalues by $\mu_\alpha\geq0$, with $\sum_\alpha\mu_\alpha=1$, one finds
\[
1-\operatorname{tr}(M_V^2)
=
\left(\sum_\alpha\mu_\alpha\right)^2
-\sum_\alpha\mu_\alpha^2
=
2\sum_{\alpha<\beta}\mu_\alpha\mu_\beta
\geq0.
\]
Hence
\[
c_4(V)\geq0
\]
for every normalized combined flavor--spatial source direction. Equality holds if and only if $M_V$ has rank one, equivalently when the source direction factorizes as $V^{\rm a}_i=v^{\rm a}e_i$. Thus factorized source directions saturate the fourth-cumulant positivity condition, whereas generic nonfactorized directions satisfy it strictly. The continuum requirement $c_4(V)\geq0$ is therefore satisfied along every combined flavor-spatial direction. This result is nontrivial because the two quartic tensor structures carry coefficients of opposite sign and are not separately non-negative. Positivity holds only for the specific holographically obtained combination $(\operatorname{tr}M)^2-\operatorname{tr}(M^2)$, which is non-negative because $M$ is a positive-semidefinite Gram matrix. As consistency checks, all pure-noise cubic couplings of \cite{SM-BuSunZhang2022} (their $\lambda_i$) carry at least one spacetime derivative, so the ultralocal cubic vertex vanishes at zero force, as parity of the seed requires; and the KMS relations among their coefficients (their Eqs. (3.79)) are the tensorial realization of the order-by-order KMS expansion \eqref{eq:tower}, rather than of the stronger force-independent-seed hierarchy \eqref{Eq:ForceIndependentRecursion}. The framework then gives a falsifiable higher-order prediction. Extending the holographic computation to sextic order in the spatial difference sources must yield
\[
c_2(V)c_6(V)\geq c_4(V)^2
\]
for every normalized combined flavor--spatial direction $V^{\rm a}_i$, where all three cumulant densities are obtained by contracting the corresponding Gaussian, quartic, and sextic tensors along the same direction. If the quadratic spatial kernel of Ref.~\cite{SM-BuSunZhang2022} is isotropic in this source space with the normalization used above, then $c_2(V)=w_2=2r_h^2/\pi$. 

Two caveats attach to this comparison. First, the mixed temporal-spatial quartic structures of \cite{SM-BuSunZhang2022} (their $\chi_{1,5},\chi_{2,5}$, involving the charge-density source $B_{av}$) can be of either sign directionally, but the associated Gaussian kernel is derivative-suppressed ($w_1=0$ at leading order), so the white-noise scaling of Sec.~\ref{sec:cont} does not apply there; this is precisely the derivative-dependent sector beyond LZDN left open in the main text. Second, the holographic action of \cite{SM-BuSunZhang2022} satisfies dynamical KMS at the quantum level, whereas the continuum test is applied here to the leading-derivative vertices treated as classical noise data. The sign and normalization in Eq.~\eqref{eq:holoS} follow from
the conventions of Ref.~\cite{SM-BuSunZhang2022}, including the
conversion between the coefficient of $b^4$ in $i S$ and the
fourth cumulant, $\kappa_4=4!\,[b^4]i S$.

\end{document}